\documentclass[prb, twocolumn]{revtex4}
\usepackage{graphicx}
\renewcommand{\[}{\begin{equation}}
\renewcommand{\]}{\end{equation}}

\begin{document}

\title{The Problem of Tori in Phase Space}

\author{Paul Stanley}

\affiliation{Department of Physics and Astronomy\\Beloit
College\\Beloit, WI 53511}

\date{September 1, 2003}

\begin{abstract}
A fundamental premise of Hamiltonian chaos is the existence and
properties of tori in phase space.  More than a geometrical construct,
these structures underlie the very dynamics of both classical and
quantal systems.  Although presented in many introductory textbooks on
nonlinear dynamics, the structure of tori in phase space is rarely as
transparent as is often presented.  Here we outline some of the
mathematics of the tori and illustrate a few of the geometric and
topologically idiosyncrasies.

\end{abstract}

\maketitle

\section{Introduction and Overview}

Motion on a torus is covered in a number of introductory texts on
non-linear dynamics.  A full understanding of why motion is on a torus
requires application of the Poincar\'{e}-Hopf theorem and, for weakly
chaotic systems, the KAM theorem.  We only provide a sparse
introduction to the notation here.  Although this approach is
inadequate for teaching, it will form the backbone for the discussion
on the physical realities of this abstract topic.  For examples of
tori in introductory texts we will rely on the resource letter
by Hilborn and Tufillaro\cite{Hil:97} which was later republished in a
collection of articles made available by the American Association of
Physics Teachers\cite{Hil:99}.

We will begin by outlining the mathematical basis for the existence of
tori in the phase space of Hamiltonian systems.  Originally presented
as abstract topological entities, we then use the action-angle
variables to provide a natural coordinate system to the tori.  This is
followed by a discussion of some of the properties of the action-angle
representation.  Finally, we examine the structure of the tori when
viewed in a common three dimensional subspace of the phase space.
Some concluding remarks are devoted to the use of symbolic math
programs to visualize the structure of phase space.

\section{The existence of tori in Hamiltonian dynamics}

Consider a point particle which can move in $n$ spatial dimensions.
The case $n=1$ effectively restricts the motion to the $x$ axis; the
case $n=2$ allows for two dimensional motion, such as that of a
projectile.  We can define the position of the object according to the
vector $\vec{q}$, which resides in an $n$-dimensional space referred to
as configuration space.  Each of the individual spatial coordinates
$q_i$ has an associated momenta coordinate $p_i$.  Both $\vec{p}$ and
$\vec{q}$ are functions of time and we can combine these two vectors
into a single, $2n$ dimensional {\em column} vector
\[
\vec{u} = \left[
\begin{array}{c}
q_1 \\ \vdots \\ q_n \\ p_1 \\ \vdots \\ p_n
\end{array}
\right]
\]
which resides in a $2n$-dimensional space referred to as phase space.
With no other constraints, the particle can exist {\em anywhere}
within the phase space.  One of the challenges is to understand the
differences between the plethora of spaces and their respective
dimensions.

For simplicity we assume that the point particle moves in a potential
$U(\vec{q})$ and has a kinetic energy quadratic in the velocities
given by $T(\vec{\dot{q}})$.  We choose the potential so that it has
explicit dependence on neither the velocity nor the time, which allows
us to write $p_i = \partial T/\partial \dot{q}_i$.  A straightforward
change of variables, combined with our previous restriction on the
form of the kinetic energy, means that the kinetic energy can be
written as $T(\vec{p})$ and then the Hamiltonian can be written as
\[
H(\vec{u}) = T(\vec{p}) + U(\vec{q}).
\]
This Hamiltonian also has no explicit time dependence.  Consequently,
\[
\partial H/\partial t = dH/dt = 0,
\]
so $H$ is a constant of the motion.  The equations of motion are given
by
\[
\label{eqn-of-motion}
d \vec{u}/dt = {\bf J} \vec{\nabla} H,
\]
where ${\bf J}$ is the $2n$ by $2n$ symplectic matrix
\[
{\bf J} = \left[
\begin{array}{rr} {\bf 0} & {\bf I} \\ -{\bf I} & {\bf 0}
\end{array} \right],
\]
and $\vec{\nabla}$ the gradient operator defined in $\vec{u}$ (phase)
space as
\[
\vec{\nabla} = \sum_{i=1}^{2n} \hat{u_i}\frac{\partial}{\partial u_i}
\]

The solution to Eq.~\ref{eqn-of-motion}, $\vec{u}(t)$, provides the
phase space coordinates of the particle as a function of time.  If the
potential $U$ is analytic, then so is the trajectory $\vec{u}(t)$,
although a closed form for the solution may not be possible.  $d
\vec{u}/dt$ is a vector in phase space which is locally tangent to the
trajectory which passes through the point $\vec{u}$.

The right hand side of Eq.~\ref{eqn-of-motion} is a unique function of
position $\vec{u}$, consequently the tangent vector is also
unique. The immediate consequence is that two trajectories cannot
cross through the same point, and a trajectory cannot cross itself.
Trajectories are necessarily a one-dimensional structure, regardless
of the dimensionality of phase space, which will fall into one of two
families: closed and open.  A closed trajectory is periodic.  An open
trajectory will fill the subspace through which it is allowed to
wander.  The dimension, geometry, and topology of this subspace is
important in discussions of chaos.

Since $H$ is a constant of motion then any trajectory is necessarily
embedded in a $2n-1$ dimensional subspace of phase space.  This
subspace is often called the energy shell.  The energy shell divides
regions of phase space into those regions with more energy and those
regions with less; consequently, the energy shell is a boundary.  The
energy shell itself, being a boundary, does not have a boundary.  When
$n=1$ the energy shell must be topologically equivalent to a circle in
a plane. Although there may be several isolated regions with the same
value for $H$, a trajectory in one region can never jump to another
region.  When $n=2$ the energy shell is three dimensional and it does
{\em not} have a two dimensional surface.

\begin{figure}[p]
\includegraphics{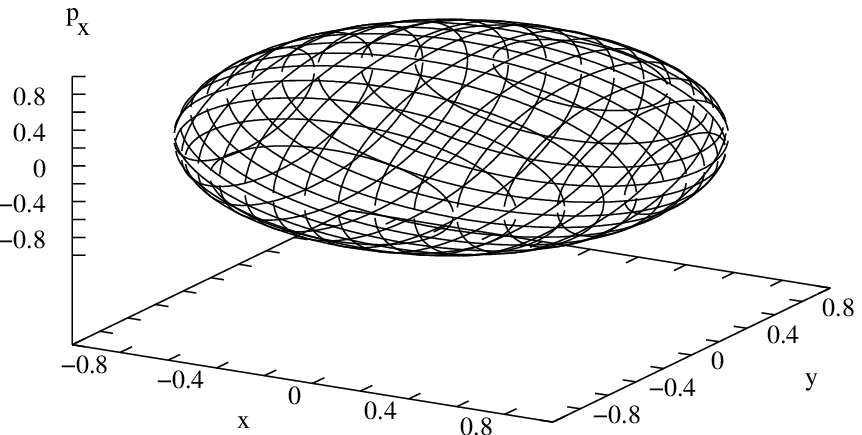}
\caption{\label{energyshell} Part of the energy shell corresponding to
$p_y\ge 0$ for two coupled oscillators.  The surface is given by
$p_y=0$, the interior of the surface is only half of the energy shell
through which the trajectory is allowed to wander.}
\end{figure}

A common way of representing the energy shell in a four dimensional
phase space is to consider only the coordinates $x$, $y$, and $p_x$.
If $H=E$ is specified, then $p_y$ is determined to within a sign by
$E$ and the variables $x$, $y$, and $p_x$.  Sketching the energy shell
requires {\em two} plots: one for $p_y \ge 0$, an example of which is
shown in Fig.~\ref{energyshell}, and one for $p_y \le 0$.  The two
plots will looks the same, but they belong in two distinct regions of
phase space and are joined on the common surface $p_y=0$.

A correct interpretation of the energy shell is to imagine a second
object much like Fig.~\ref{energyshell} which would correspond to
$p_y \le 0$.  Turn this second object inside-out, so that it now
occupies all of the space which is {\em not} occupied by the region
$p_y\ge 0$.  Then glue the two regions along the free surface.  The
resultant energy shell will have {\em no} surfaces.

It is possible for the equations of motion to have, in addition to the
Hamiltonian, linearly independent conserved quantities $F_i(\vec{u})$
which are explicitly functions of only the phase space coordinates.
These quantities are called integrals of motion; the condition of
linear independence is supplemented by the requirement of the
vanishing of the Poisson bracket:
\[
\label{involution}
0 = \left[F_i, F_j\right] = \sum_{k=1}^n \left( \frac{\partial
  F_i}{\partial q_k} \frac{\partial F_j}{\partial p_k} -
  \frac{\partial F_i}{\partial p_k} \frac{\partial F_j}{\partial q_k}
  \right)
\]
where we choose, for convenience, the first integral of motion to be
the Hamiltonian: $F_1 = H$.  The condition that Eq.~\ref{involution}
vanishes for all $F_j$ if $F_1 = H$ is readily seen since
\[
\frac{dF_i}{dt} = \left[F_i, H \right] + \frac{\partial F_i}{\partial
  t}.
\]
If $F_i$ is a constant of motion the left side vanishes; if $F_i$ is
not explicitly a function of time the partial derivative vanishes.

A system is in involution, or completely integrable, if there exists
$n$ linearly independent integrals of motion, one for each of the
dimensions in coordinate space $\vec{q}$, and these integrals of
motion satisfy the Poisson bracket conditions of Eq.~\ref{involution}.
Such a system is ripe for quantization, as the Poisson bracket
formulation is the classical analog of commutation in quantum
mechanics and the integrals of motion correspond to the simultaneous
observables.

For a one dimensional configuration space there is one integral of
motion which can be chosen to be the Hamiltonian.  Other choices are
possible.  For a two dimensional configuration space there are two
integrals of motion, one of which could be the Hamiltonian; the other
might be, for example, the angular momentum.

Each constant of motion $F_i(\vec{u})$ restricts the allowed space for
the trajectory $\vec{u}(t)$.  $n$ integrals of motion will restrict
the trajectory to reside in an $2n-n=n$ dimensional subspace of phase
space. It is the topology of the restricted subspace that is of
interest here.

Each integral of motion has a corresponding vector flow field,
\[
\label{flow-field}
\vec{\xi_i} = {\bf J} \vec{\nabla} F_i,
\]
which should be compared to Eq.~\ref{eqn-of-motion}, for which
$H=F_1$.  The vector field $\vec{\xi_i}$ is necessarily tangent to the
surface defined by $F_I(\vec{u}) =$ constant.  In addition, except for
a set of measure zero, $\vec{\xi_i}(\vec{u})$ never vanishes.

Note that we can write the Poisson bracket in another form:
\[
\sum_{k=1}^n \left( \frac{\partial F_i}{\partial q_k} \frac{\partial
  F_j}{\partial p_k} - \frac{\partial F_i}{\partial p_k}
  \frac{\partial F_j}{\partial q_k} \right) = \left(\vec{\nabla}
  F_i\right)^{\sf T} {\bf J} \vec{\nabla} F_j.
\]
This can be combined with Eq.~\ref{flow-field} to yield
\[
\left[ F_i, F_j \right] = \left({\bf J}\vec{\nabla}
  F_i\right)^{\sf T} {\bf J} {\bf J}\vec{\nabla} F_j =
  \vec{\xi_i}^{\sf T} {\bf J} \vec{\xi_j}.
\]
Since the Poisson bracket vanishes for the integrals of motion, the
quantity $\vec{\xi_i}^{\sf T} {\bf J} \vec{\xi_j}$ also vanishes.  One
should consider this the symplectic equivalence of orthogonality.

The trajectory, then, must lie in an $n$ dimensional subspace of a
$2n$ dimensional phase space.  This $n$ dimensional subspace must also
contain $n$ linearly independent non-vanishing vector fields.  It is
from the Poincar\'{e}-Hopf theorem that we deduce that the subspace
must be topologically equivalent to a torus.  Our argument here is
similar to Tabor\cite{Tab:89}, but a readily accessible, fuller
argument can be found in the text by Nash and Sen\cite{Nas:83}.  We
say that these are invariant tori because any trajectory which
originates anywhere on the tori will remain on the tori for all time.

It is worthwhile to provide a definition of the torus, since many
students will, upon hearing the word torus, immediately imagine the
interior of a donut or bagel.  A 1-dimensional torus $T_1$ is a line
segment with the endpoints joined together: in effect, a closed loop.
The torus, in this case, is the set of points which make up the
circumference.  The torus $T_1$ is homomorphic to the sphere $S_1$.
In the 1-dimensional case the torus is the surface defined by $H=$
constant.  It is not necessary for $T_1$ to look identical to a
circle; any closed single loop, however deformed, would qualify as $T_1$.

\begin{figure}[p]
\includegraphics{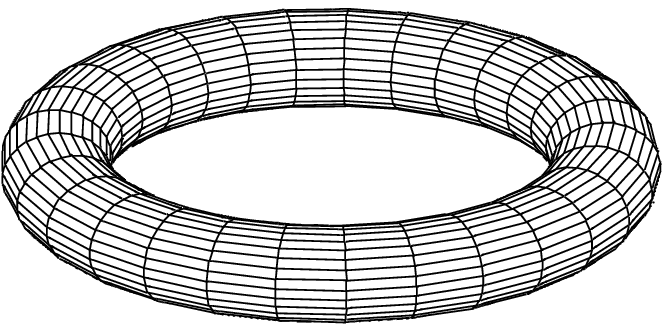}
\caption{\label{T2} A representation of $T_2$ in a two dimensional
projection of a three dimensional space.  The lines are not physical
trajectories.}
\end{figure}

A 2-dimensional torus $T_2$ is a flat sheet with opposite sides first
joined together to make a tube, and then the tube ends joined together
to make a donut shaped object.  The torus in this case is the sheet,
not the interior.  The torus can also be written as $T_2 = S_1 \times
S_1$.  Figure~\ref{T2} is a representation of the torus $T_2$.  The
lines in the figure are not representative of a physical trajectory;
all that is known at this point is that a physical trajectory must lie
on the surface of the torus and that the trajectory lines must not
cross each other.  The regularity of the figure is potentially
misleading: the torus does not need to be round or symmetric.

A 3-dimensional torus would be constructed in much the same way: $T_3
= S_1 \times S_1 \times S_1$.  It is not possible to construct a
meaningful sketch of $T_3$ on a two dimensional surface, nor is it
easy for a student to visualize.

It is the application of the Poincar\'{e}-Hopf theorem that is most
often glossed over in introductory text books.  The existence of a
{\em single} non-vanishing tangent vector field on a closed surface
restricts the allowable type of surface.  Since the non-vanishing
tangent vector field is analogous to combing the hair on a surface,
this problem is often known as the hairy ball theorem, which is {\em
not} the same as the Poincar\'{e}-Hopf theorem.  It is true that the
torus $T_n$ satisfies the combability requirement for a single vector
field, but so can an odd-dimensional sphere $S_{2n-1}$. A one
dimensional sphere $S_1$ is topologically equivalent to a circle; a
two dimensional sphere $S_2$ is the traditional sphere; a three
dimensional sphere $S_3$ would be embedded in a four dimensional
space.

$S_1$ and $T_1$ are equivalent, while $S_2$ is not allowed through the
hair combing argument.  For one and two dimensional configuration
spaces, such as would be found for a single oscillator or two coupled
oscillators, the hair combing argument is sufficient to restrict the
motion to a torus.  But for higher dimensional systems, one must
return to the Poincar\'{e}-Hopf theorem and refer to the existence of
$n$ linearly independent vector fields.  

At this point the tori are physical but still highly abstract
structures.  That they exist is important for both chaos theory and
WKB quantization (for examples on both, see Reichl\cite{Rei:92}).  The
tori must exist regardless of the choice of coordinates for phase
space (for example, in both ${x,y,p_x,p_y}$ and ${r,\theta, p_r,
p_\theta}$).  There is, however, a natural coordinate system to use
when considering the tori.

\section{Action angle variables}

Finding the integrals of motion is one of the challenges of classical
mechanics.  In a few, but still important, cases, the Hamiltonian is
separable, and one can use the action-angle coordinates to find a set
of integrals of motion.

The expression for the Hamiltonian can be inverted to obtain the
momentum in terms of position and energy, $p(q,E)$.  For a one
dimensional periodic system the action is defined as
\[
\label{1D-action}
I(E) = \frac{1}{2\pi} \oint p(q,E)\, dq,
\]
where the integral is taken over one period\footnote{Not all authors
include the factor of $1/2\pi$, but they should.}.  For the one
dimensional case the path of integration is necessarily a path for a
physical trajectory.  Upon integrating over a complete period the
action can be found as a function of energy $I(E)$. It is then
possible to find the energy, and hence the Hamiltonian, as a function
of the action $H(I)$.  Once the transformation is complete $H$ is {\em
only} a function of the action.

The action, $I$, is a generalized momentum.  The corresponding
position variable is an angle $\theta$; the equations of motion are
now
\begin{eqnarray}
\dot{\theta} &=& \partial H(I) / \partial I = \omega,\\
\dot{I} &=& \partial H(I) / \partial \theta = 0.
\end{eqnarray}
Consequently $I$ is a constant of motion and $\omega$ is the angular
frequency of the periodic motion through $\theta$ space.  Since
$\theta$ and $I$ form a set of generalized position and momentum
variables, we should be able to use them in Eq.~\ref{1D-action}, or
\[
I = \frac{1}{2\pi} \oint I(E)\, d\theta,
\]
where it is apparent that $\Delta \theta = 2\pi$ over a single period.

\begin{figure}[p]
\includegraphics{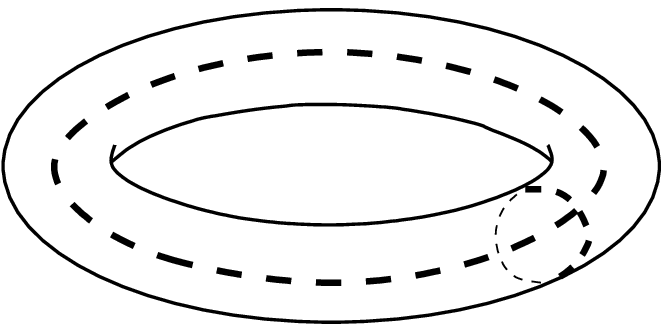}
\caption{\label{2Dwinding} Two distinct paths of integration for the
torus $T_2$.}
\end{figure}

That the path of integration is a single winding around a torus $T_1$
is a hint of the technique for higher dimensional systems.  In the two
dimensional system we do not integrate along the path of a physical
trajectory; we instead integrate along topologically distinct paths.
We restrict consideration to paths with a winding number of one.  For
a two dimensional system there are two possible paths, as shown in
Fig.~\ref{2Dwinding}.  These are distinct in that it is not possible to
deform one path into the other while staying on the torus.  Neither
path is null in that it is not possible to shrink either path to a
point without leaving the surface of the torus.  For a torus $T_n$
there are $n$ topologically distinct paths; none of which are likely
to belong to a physical trajectory.

The action variables in the $n$ dimensional configuration space are
therefore defined similarly to the one dimensional case:
\[
\label{generalized-action}
I_i = \frac{1}{2\pi} \oint_{C_i} \vec{p} \cdot d\vec{q}
\]
where the paths $C_i$ are {\em any} $n$ topologically distinct paths
on the tori $T_n$.  It is not important where we choose the paths
$C_i$, as the integral $I_i$ is the same for all paths which are
topologically equivalent.  Hence, for the torus $T_2$, there are only
three possible values for the integral in
Eq.~\ref{generalized-action}, one of which is $I=0$.  

For the remainder of this paper we will restrict ourselves to a two
dimensional configuration space, a three dimensional energy shell, and
four dimensional phase space.

\begin{figure}[p]
\includegraphics{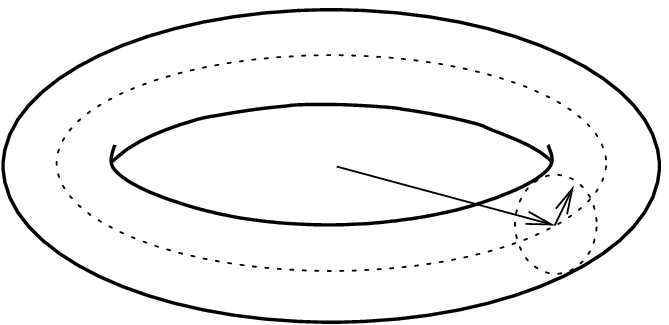}
\caption{\label{actionangle} The action-angle representation.  The two
arrows represent the action variables $I_1$ and $I_2$.}
\end{figure}

The use of the action-angle variables provide a convenient coordinate
system for the tori.  Any torus can be uniquely specified by the
values of the actions $I_1$ and $I_2$.  From a purely geometric point
of view the actions are the two radii of the torus, as is shown in
Fig.~\ref{actionangle}.  Any point on the torus is specified by the
value of the angle variables $\theta_1$ and $\theta_2$.  A trajectory
on the torus will eventually cover all points on the torus if the
ratio $\omega_1/\omega_2$ is irrational; if not, the trajectory
occupies a one dimensional subspace of the torus.  The former type of
orbit is called quasi-periodic, the latter is periodic.  The tori
occupied by these two types of orbits are referred to as irrational and
rational tori, respectively.

Although a common representation, which is used in many texts, it is
not as straightforward as might be hoped.  Since few texts show
pictures of more than one torus at a time, the student is left with
the challenge of trying to decipher (1) if there is or is not more
than one torus and (2) how these tori might arrange themselves in
phase space.

Clearly there is more than one tori.  Every point within the energy
shell in phase space must lie on a trajectory, and every trajectory,
to within a set of measure zero, is embedded in a subspace which has
the topology of a torus.  It follows that the energy shell must be
filled with an infinite number of tori.

\begin{figure}[p]
\includegraphics{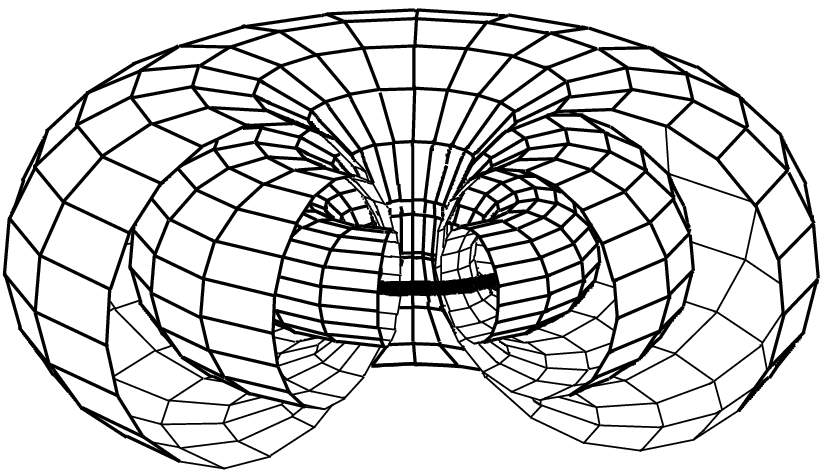}
\caption{\label{nestedtori}A representation for the layered tori for
  any separable Hamiltonian which can be written as $H(I_1,I_2) =
  f_1(I_1) + f_2(I_2)$.}
\end{figure}

By considering the limitations on the individual trajectories which
make up the tori we can deduce that the tori are necessarily nested,
cannot cross each other, and must fill all of the energy shell, except
for regions of measure zero.  The only possible nesting for a pair of
coupled oscillators would be topologically equivalent to
Fig.~\ref{nestedtori}.  In fact, any bound separable two dimensional
system must have nested tori with a topology equivalent to
Fig.~\ref{nestedtori}.

Part of the confusion that invariably arises is due to the implicit
relationship between the action variables.  Within an energy shell the
action variables are not independent; they must satisfy some
relationship of the form
\[
H(I_1,I_2) = f_1(I_1) + f_2(I_2)
\]
for a separable system.  This means that there is both a maximum and a
minimum value for both $I_1$ and $I_2$.  The thick black ring in
Fig.~\ref{nestedtori} could represent the torus with $I_1$ a maximum
and $I_2$ a minimum; here $I_2=0$ and the torus is really a one
dimensional ring.  The torus corresponding to $I_1=0$ is not shown in
the figure.  It would be a vertical line through the center which
extended to infinity in both directions.  Positive infinity and
negative infinity would be joined to complete the loop.

It must be clarified that changing from traditional position and
momentum variables to action-angle variables will not reduce the
dimensionality of the system, nor will it simplify the topology.  It
will simplify the representation in a figure, and serve as a useful
tool for labeling the tori.

\section{Finding the tori}

The easiest pictorial representation of the one dimensional torus,
$T_1$, is on the phase plot.  For a 2-dimensional phase space the tori
are 1-dimensional and all tori will appear as generalized circles.

\begin{figure}[p]
\includegraphics{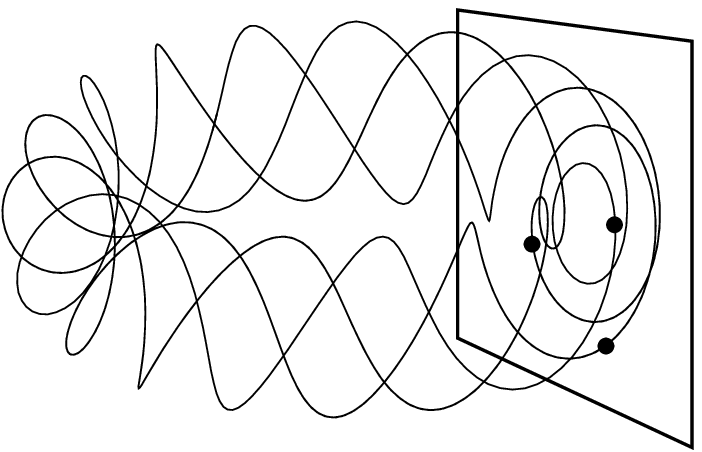}
\caption{\label{sos}The construction of the Poincar\'{e} surface of
  section.  The trajectory winds around the torus; points are
  plotted where it crosses a specified plane.}
\end{figure}

The traditional representation of higher dimensional tori is through
the use of the Poincar\'{e} Surface of Section.  As a trajectory wanders
through phase space according to $\vec{u}(t)$, we plot the position
every time the trajectory crosses through a specified plane.  The plane
is traditionally taken to be the $x,p_x$ plane where $y=0$.  The
fourth phase space variable, $p_y$, is determined by energy
conservation.  In order to avoid ambiguity the position is only plotted
when $p_y>0$.  Graphically, this would look similar to Fig.~\ref{sos}

Figure~\ref{sos}, or the equivalent, is common in introductory texts
which deal with tori; it, or something similar, can be found in the
books by Moon\cite{Moo:92}; Berg\'{e}, Pomeau, and Vidal\cite{Ber:84};
Abraham and Shaw\cite{Abr:92}; Tufillaro, Abbott, and
Reilly\cite{Tuf:92}; Seydel\cite{Sey:88} and others.  A similar
picture also appears in the text by Hilborn\cite{Hil:94},
except that in the Hilborn text a third axis, $y$ is drawn which has
no physical reality in the space occupied by the torus.

If the trajectory is allowed to run around the torus for a
sufficiently long time the points of intersection with the surface
will eventually make a closed circular loop.  These diagrams are quite
common in introductory texts, although rarely is an effort made to
identify the points on a surface of section with the physical
trajectories through either phase space or configuration space.  A
number of tori intersecting with the surface of section would appear
as loops, and are often represented by concentric rings.  However,
there is no need for symmetry, and the intersection of the tori with
the surface of section can produced rather tortured loops with bizarre
tendrils.

The construction of the Poincar\'{e} surface of section as described
above is valid, but the graphics which are so often used to describe
the construction are misleading.  After learning that the intersection
will occur on the $x, p_x$ plane where $y=0$, it is tempting to assume
that the torus will actually possess a donut shape in some three
dimensional space such as ${x, p_y, y}$.

Part of the challenge arises from the portrayal of the energy shell as
a three dimensional solid with a surface.  Figure~\ref{energyshell}
only shows the {\em half} of the energy shell where $p_y>0$, and any
trajectory which crosses the $x,p_x$ plane where $y=0$ will sooner or
later need to cross it again except going in the other direction.
Trajectories spend only half of the time in the energy shell where
$p_y>0$.  If a trajectory is plotted in a three dimensional
space ${x, p_y, y}$ where $p_y>0$, then it will only show half of the
torus.

\begin{figure}[p]
\includegraphics{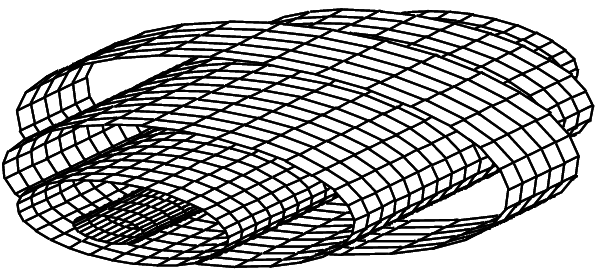}
\caption{\label{toriphase3}The appearance of nested tori in a three
  dimensional ${x,p_x,y}$ slice of phase space.}
\end{figure}

Figure~\ref{toriphase3} is an illustrating of the structure of the
nested tori of coupled harmonic oscillators when viewed in the three
dimensional space of ${x, p_x, y}$ with $p_y>0$.  In the figure $p_x$
is the vertical axis; it is easy to imagine that the intersection with
the $x,p_x$ plane will produce concentric rings.  The figure should be
compared with Fig.~\ref{energyshell}, as both are concerned with the
same Hamiltonian.  In fact, the surface of Fig.~\ref{energyshell} is
an envelope that contains the half-tori of the figure.

How does Fig.~\ref{toriphase3} compare to Fig.~\ref{nestedtori}?
There are two ways to map one figure onto the other; the easier method
is to rotate Fig.~\ref{toriphase3} through $90^{\circ}$.  Then
Fig.~\ref{toriphase3} corresponds to the very center of
Fig.~\ref{nestedtori}.  The outermost ``ring'' in
Fig.~\ref{toriphase3} is the dark ring through the center of the
nested tori of Fig.~\ref{nestedtori}.

\section{Concluding remarks}

A symbolic math program such as {\em Maple} or {\em Mathematica} can
be used to quickly illustrate the three dimensional structure of {\em
part} of the actually tori.  This structure can be projected onto
slices of the phase space so that the values of the action can be
quickly calculated, and from this the frequency ratio of the tori can
be computed.  Unfortunately such a technique is too crude to ascertain
whether the torus is a rational or irrational structure.  In addition,
the tori which are constructed by the differential equation packages in
these programs often fail to distinguish between regions of $p_y>0$
and $p_y<0$; the resulting graphical images are therefore incorrect
representations of the tori in phase space.  Still a useful tool for
visualization, the programs must be used with care when interpreting
the topology of phase space.

The importance of visualization should not be understated.  Since the
tori of phase space are so crucial to the arguments of both classical
chaos and quantum mechanics, it is imperative that a student
understand both the geometry and the topology of these structures.
Introductory textbooks, for the most part, present tori correctly; it
is, however, easy for the student to mislead themselves when trying to
develop a mental picture for the topology of the tori.

\bibliography{StanleyAJP}

\end{document}